# Mono-stability of sharp tips interacting with surface hydration layers

Victor Barcons[1], Sergio Santos[2,3], William Bonass[3], Josep Font[1], Neil H Thomson[2,3]

[1]Departament de Disseny i Programació de Sistemes Electrònics, UPC - Universitat Politècnica de Catalunya Av. Bases, 61, 08242 Manresa, Spain [2]School of Physics and Astronomy, University of Leeds, LS2 9JT, UK [3]Department of Oral Biology, Leeds Dental Institute, University of Leeds, LS2 9LU, UK

[3]Laboratory for Energy and NanoScience (LENS), Institute Center for Future Energy (iFES), Masdar Institute of Science and Technology, Abu Dhabi, UAE

Abstract

It is generally thought that capillary interactions in nanoscale contacts give rise to unwanted behaviour due to high adhesion. We show that this is not the case for sufficiently small contacts in ambient conditions. High resolution ambient atomic force microscopy AFM requires tip-sharpness, proximity and small forces, but the cantilever dynamics might not allow these three conditions to be met simultaneously. Hitherto, accepted dogma is that small drive amplitudes lead to either tip trapping or L mode (attractive) imaging, where proximity is inhibited. Here we show that the hydration layer might be responsible for allowing the AFM tip to be brought stably within angstroms of the surface using a small amplitude small set-point (SASS) mode. This phenomenon enhances resolution and stability while dramatically reducing tip wear.

Surface hydration is a common phenomenon in ambient conditions. Layers of water of nm and sub-nm thickness are responsible for a myriad of phenomena with macroscale implications such as adhesion and cohesion, lubrication and tribology [1-4]. Moreover, in the nanoscale, water can give rise to phenomena with no precedents in the macroscale due to confinement and limited spatial molecular motility [5,6]. In the field of nanotechnology, modelling and predictions have complemented instrumentation and experimental outcomes when understanding the behaviour of water in the nanoscale since the beginning [2,3,7,8]. In this respect, the AFM allows both researching nanoscale water interactions[2] and develop nanofabrication techniques where hydration is involved [9,10]. Adhesion and water interactions have also played a fundamental role in the development of the AFM.

Dynamic AFM (dAFM) modes were developed [11,12] soon after the invention of the AFM[13] and have been essential for the progression of biological imaging[14-16]. Oscillating the cantilever eliminates damaging shear forces that arise in contact mode[11,12,14,15,17]. Small oscillation amplitudes ($A_{sp}$<1-5nm) have long been thought[11,12,18,19] to lead to tip trapping due to capillary and/or van der Waals (vdW) adhesion forces[12,19,20]. For example, the stability criterion[21,22] has been used to establish that either large oscillation amplitudes, stiff cantilevers or both are required to avoid tip-trapping and/or to minimize background noise due to tip trapping. On one hand, the combination of larger amplitudes and very stiff cantilevers (bringing about high force sensitivity and stability when the tip is close to the sample), led Frequency

Modulation (FM) to be the first AFM technique to achieve atomic resolution under ultra high vacuum conditions[22-24]. On the other, small oscillation amplitudes $A_{sp}$ imaging[25-28] is now thought to be essential for advances in high resolution even under liquid[29-31]. Nevertheless, the stability criterion establishes [21, 22] that stability should decrease with decreasing oscillation amplitude as a result of energy imbalances; the energy being dissipated increases relative to the cantilever's stored energy with decreasing oscillation amplitude.

In this letter, we investigate the dynamics of ultra sharp tips at angstrom and nanometer equilibrium tip-surface separations $z_c$ (Fig. 1) in ambient conditions for ultra small amplitudes. We refer to the oscillation amplitude as $A_{sp}$ when feedback is on and as A when off. The study focuses on both small driving or free amplitudes and oscillation amplitudes: these being two required conditions for high resolution. We further focus the study on ultra sharp tips (but otherwise standard probes) with radii of radii R≤5nm and standard values of cantilever stiffness in ambient conditions, i.e. k>2-10N/m. In particular, we have used k=40 N/m here throughout. These are required conditions for the observation of the phenomena here investigated; R controls the character and magnitude of the non-linearities in the tip-surface interaction[15] while k controls the mean energy stored in the cantilever and provides stability[21, 22]. That is, we have observed this phenomena both experimentally and in simulations for relatively high values of stiffness, as above mentioned, and whenever using tips of R<5nm as confirmed by Scanning Electron Microscope (SEM) scans (Fig. 1d). Contrary to accepted dogma, we show that the water layers are responsible for a phenomenon that 1) controllably and smoothly drives the cantilever at very close proximity to the surface with ultra small oscillation and drive amplitudes and 2) makes the system there monostable, thus greatly increasing the stability of the system under these conditions. Thus, the phenomenon deals with two of the great challenges of high resolution AFM and, significantly, occurs when using cantilevers of standard stiffness in ambient conditions. We use a point-mass model to simulate the cantilever dynamics. For our set-up and simulations higher modes can be neglected [32, 33]; the point-mass model is a good approximation to the real phenomenon[34]. The equation of motion has been implemented in C and solved numerically with fourth and eighth order Runge-Kutta algorithms. We term the net tip-surface force in the interaction $F_{ts}$ (Fig. 1). Furthermore, each of the distance dependent components comprising $F_{ts}$ that we use here is defined below and differentiated with the use of specific subscripts. Smooth here refers to transitions between oscillation states which do not display discrete steps in amplitude. [35]

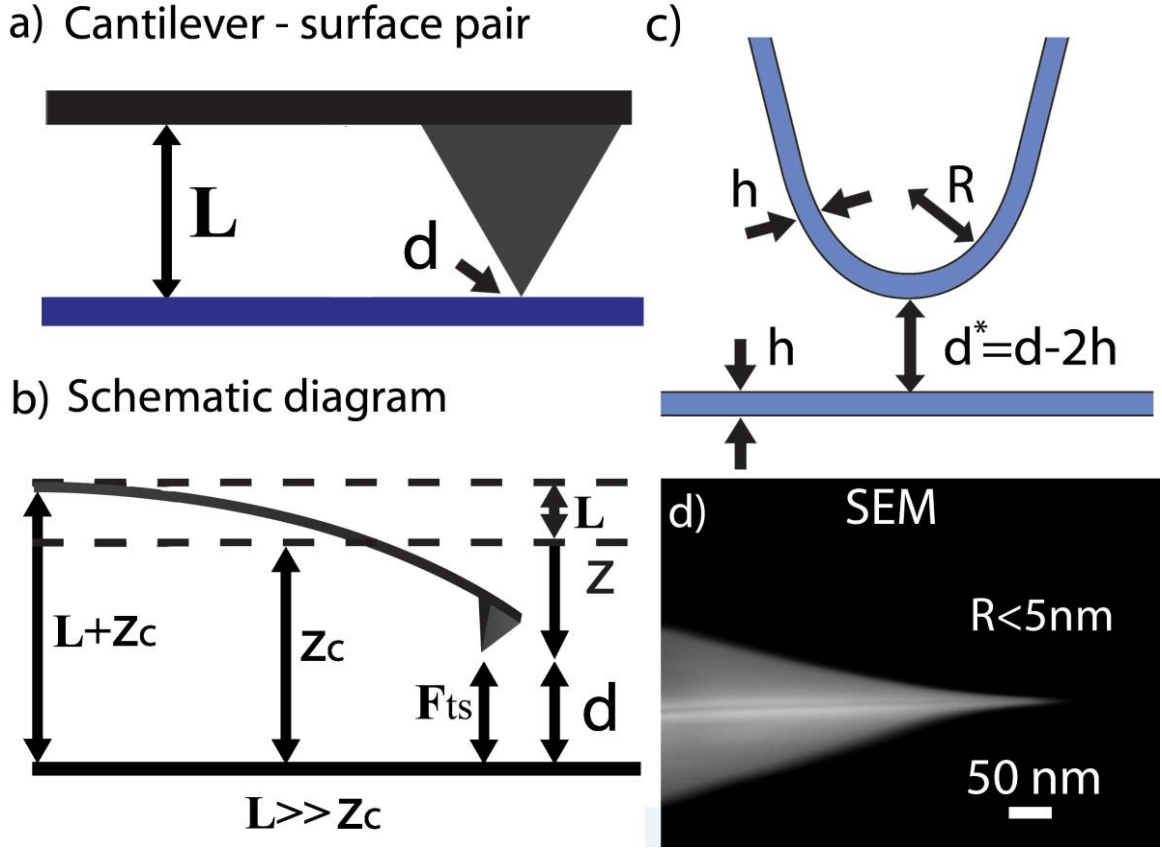

FIG. 1: (a) Schematic representation of a cantilever probe interacting with a surface. L is the length of the probe in the vertical axis relative to the cantilever and it is typically of micro-scale dimensions. Thus this distance is never used as a reference. (b) Schematic of the useful distances and separations for the tip interacting with the surface. Here $z_c$ is the equilibrium tip-surface separation for the unperturbed cantilever where $z_c \ll L$. Moreover, z is the instantaneous position of the tip relative to $z_c$ and d is the instantaneous tip-surface distance. (c) Schematic representation of the end of the tip with effective radii R when both the tip and the surface are hydrated and covered by a layer of water of height h. The effective distance between the water on the tip and the water on the surface is $d^*$ where $d^* = d-2h$. (d) SEM image of the end of one of the cantilevers displaying the phenomena described in this article. Only those cantilevers for which R< 5nm display the phenomena detailed here, i.e. N and SASS regions as observed in Figs. 2a (experimental) and 2c (simulation). Adapted from PhD thesis of S Santos[36] and from ref 37. [37]

The Derjaguin-Muller-Toporov (DMT)[38] contact forces and the long range van der Waals[39] forces have been used to model the conservative part of the interaction as before[34, 40, 41]; this is a suitable choice for sharp tips[34, 42, 43]. We use the capillary force and an effective interaction distance between the tip and the surface $d^*$ (Fig. 1c). That is, the interaction at larger separations occurs between the water layer on the tip and surface respectively (see supplementary for details). This addition is a key factor to understand small oscillations in ambient AFM (Figs. 2-4). The tip-surface distance (d) dependencies are then as follows

$$F_{ts}(d^*) = -\frac{H_{H_2O}R}{6(d^*)^2} \qquad (1)$$

when $d > d_{off}$ and also when $d_{on} < d < d_{off}$ provided the capillary force is not acting[40]. H is the Hamaker constant. The effective distance, $d^*$, is defined as $d^* = d - 2h$, where h is the height of the water layer on the tip and surface respectively (Fig. 1c). The suffixes on and off make reference to the distance for which the capillary is formed as the tip approaches the surface and the distance when it breaks during retraction respectively[40]. Here $a_0$ is an intermolecular distance[41] typically taken to be 0.165nm[4]. When the capillary is on and provided $d > a_0$ the tip-surface force is

$$F_{ts}(d) = F_a(H^*) + F_{CAP}(d) \qquad (2)$$

where[44]

$$F_{CAP}(d) = -\frac{2\pi\gamma_{H_2O}R}{1 + \frac{\pi R d^2}{V_{men}}} \qquad (3)$$

and

$$F_a(H^*) = -\frac{H^* R}{6a_0^2} \tag{4}$$

where $F_{CAP}$ is the capillary force and $F_a$ is the adhesion force to the vdW interactions. $V_{men}$ is the volume of the meniscus derived from geometrical considerations.[40] Since the water layers are in contact when the capillary bridge forms, $F_a$ is saturated here in terms of distance. H is nevertheless interpolated when the capillary neck is formed since the interaction changes from a water-water interaction, i.e. water on the tip and water on the surface, to a tip-surface interaction. Thus we write

$$H^* = \frac{\Delta H}{(d_{0ff} - a_0)} d + H_S - \frac{\Delta H a_0}{(d_{0ff} - a_0)} \tag{5}$$

where $\Delta H = H_{H_2O} - H_s$ and $H_{H_2O}$ and $H_s$ stand for the Hamaker constant for the tip-water and the tip-surface respectively.

For the contact region, $d < a_0$ and the force is

$$F_{ts}(d) = F_a(H_s) + F_{CAP}(d) + F_{DMT}(d) \tag{6}$$

where[38]

$$F_{DMT}(d) = \frac{4}{3} E^* \sqrt{R(a_0 - d)^3} \tag{7}$$

where $E^*$ is the effective elastic modulus of the tip-surface pair as typically used in contact mechanics.[45]

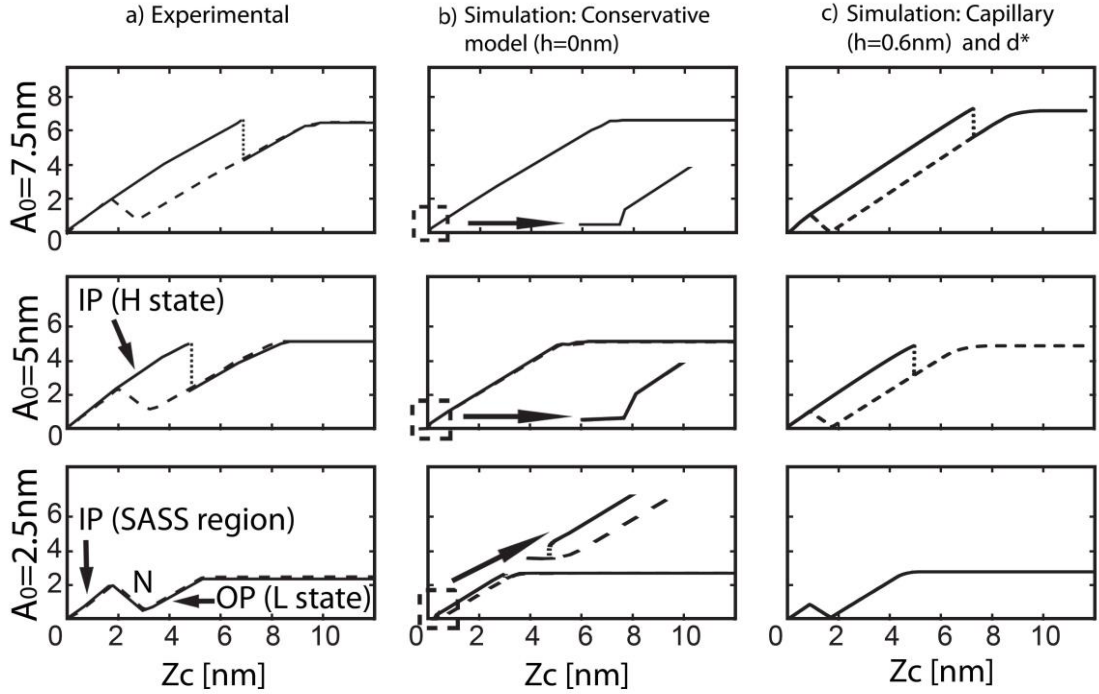

Fig 2: (a) Experimental AD curves for $A_0$=2.5, 5 and 7.5nm. The vertical axis shows A as a function of $z_c$. The amplitude reduction is approximately linear with decreasing $z_c$ where positive slopes are observed. The experimental parameters are: $f_0$=312kHz (resonant frequency), $f=f_0$ (driving frequency), k~40N/m, Q=550 (Q factor), RH=40% and R<5nm. Simulations for the same range of $A_0$ where (b) only the conservative potential has been used and (c) the capillary force and $d^*$ has been incorporated to the model. The arrows point to zoomed views for the smaller values of $z_c$. Simulation parameters: (b) $f=f_0$=300kHz, k=40N/m, R=2.5nm, Q=500, $\gamma$=40mJ (surface energy of the surface), E=10GPa (elastic modulus of the surface[46]), $E_t$=120GPa (elastic modulus of the tip); (c) as above plus h=0.6nm[47] and $\gamma_{H2O}$=72mJ (surface energy of water).

In Fig. 2(a) we show experimental amplitude-distance (AD) curves taken on a mica surface at 40% relative humidity (RH) for small $A_0$. The linear relationship $\Delta A/\Delta z_c$~1 applies in the regions of positive slope. For smaller $A_0$ (Fig. 2(a) bottom; $A_0$=2.5nm), and as the cantilever approaches the surface (dashed lines) there is an outer region of positive (OP) slope in A. Then a region of negative slope (N) follows. Finally an inner region of positive slope (IP) is observed. This IP region is approximately 3-4nm closer to the surface than the OP region thus implying more proximity, in the order of nm, in the former. The same path is followed during retraction (continuous lines) with no hysteresis. In the OP region, the mode of oscillation is typically termed L-state of oscillation and generally leads to non-contact (nc) imaging for these small values of $A_0$[41, 48]. As $A_0$ increases, the cantilever stays in the IP region during retraction (Fig. 2(a) middle and top). The IP region is in fact the H-state of oscillation where intermittent contact generally occurs[15, 34]. The trajectory followed when approaching the surface is always the same; OP, N and IP for all $A_0$. This is a general characteristic of an ultra sharp tip in ambient conditions for $A_0<2$-3nm. The experimental ADs can be compared with simulations in Fig. 2(b) where only the conservative potential has been used (h=0). The conservative potential alone does not reproduce the experimental behaviour. In Fig. 2(c), the capillary forces and $d^*$ have been added for h=0.6nm and the N and IP regions are reproduced. The existence of the N and IP regions is of significant experimental relevance for several reasons. Physically, it implies that the IP region can be reached smoothly through the N region and the tip can get to within angstrom/s of proximity for small values of $A_0$ there (i.e. $A_0<2$-3nm and $A_{sp}<0.1$-0.2nm). This can be done without the occurrence of step-like attractive (L state) to repulsive (H state) transitions[49] that can potentially lead to dramatic peak tip-surface forces[35]. Furthermore, the IP region past the local maxima in A (bottom part of Fig. 2a) is monostable, that is, only the H state exists there. Thus, because of these physical differences, we refer to the IP region past the local maxima in A as the SASS region and to the IP region before it as the standard H state. A simulated AD curve where the regions are differentiated is shown in Fig. 3a. The SASS region gives name to the high resolution low tip-wear mode of operation here presented. We have observed the SASS and N regions at both relatively low, i.e. ~10%, and very high, ~95%, values of relative humidity (data not shown). This is not surprising since the meniscus can be observed even at zero relative humidity [50] and is one of the main forces acting in ambient conditions.[19]

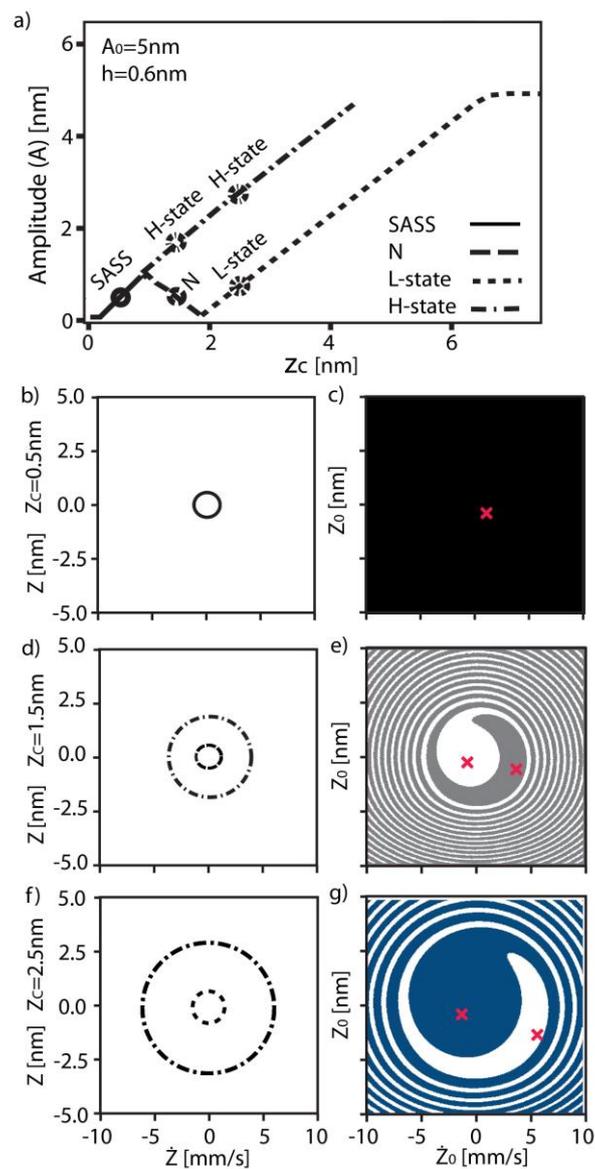

Fig 3: Simulations. (a) Simulated AD curve, where the L and H states and the N and SASS regions are reproduced. The markers indicate discrete values for which the limit cycles and basins of attraction have been calculated. A switch between two states can take place wherever there are two solutions for a given $z_c$. In the left column (b,c,d), the limit cycles (z, ż) for $z_c$= 0.5, 1.5 and 2.5nm are shown. The vertical axis is the instantaneous tip position, z and the horizontal axis is the instantaneous tip velocity, ż in the steady states. Where there are two ellipses, two limit cycles co-exist. In the right column (c, e, g), the respective basins of attraction ($z_0$, $ż_0$, $t_0$=0) are shown. These are coloured in black (SASS), grey (N region), blue (L-state) and white (H-state). The Poincaré sections are marked with red crosses. The parameters are the same as those for Fig 2c with $A_0$=5nm.

The monostability of the system in SASS can be verified with the use of phase space diagrams as shown in Figs. 3(b-g). In Fig. 3a a simulated AD curve shows the separations $z_c$ at which the phase space diagrams have been acquired. On the left column, the limit cycles are shown for $z_c$=0.5nm, $z_c$=1.5nm and $z_c$=2.5nm where the vertical axes are instantaneous position z and the horizontal instantaneous velocity ż. For $z_c$=0.5nm (Fig. 3b) there is a unique limit cycle. This limit cycle forms a closed loop with an amplitude of approximately 0.5 nm implying that the minimum separation distance here is of angstroms or fractions of an angstrom. Moreover, this further implies that the tip in this region is never more than 1nm farther away from the surface. This has been verified in the simulations (data not shown). The corresponding basins of attraction for $t_0$=0 are shown in Fig. 3c. Here the set of initial conditions ($z_0$, $ż_0$, $t_0$=0) leads to a single attractor, which, accordingly, leaves a black square (the SASS region) which physically implies monostability. The Poincaré section (z, ż) for $t_0$=0 is shown with a cross. When $z_c$ is increased to 1.5nm (Figs. 3d-e), two limit cycles are observed corresponding to the N region (grey) and the H state (white) respectively. Further increasing $z_c$ to 2.5nm (Figs. 3f-g) results in the standard L and H states as limit cycles. For each limit cycle there is an ellipse (left column in Fig. 3). Since the monostable and intermittent contact (SASS) region can be reached with very small values of $A_0$, the average and peak forces can be greatly reduced relative to the standard H state.[51] For example, note that the SASS region is already observable for $A_0$= 2.5nm in Fig. 2a even with oscillation amplitudes in the order of angstroms, i.e. $A_{sp}$~1Å. One can appreciate the relevance of this phenomenon by observing that 1) forces rapidly scale with $A_0$[41, 51] and 2) the effective area of interaction and the tip radii R also do[51].

In Fig 4a, a simulation shows how the effective radius of interaction $<r>$ varies with decreasing separation $z_c$. The way in which we have calculated $<r>$ is described in detail in the supplementary. But, briefly, $<r>$ stands for the radius of interaction in the dynamic form of AFM.[51] For simplicity and to allow comparison, the parameters are the same as those used in Fig. 3. Observe how $<r>$ forms plateaus in the L and H-states, while in the N region, $<r>$ monotonically decreases. Finally $<r>$ monotonically decreases in the SASS region and reaches a minimum there (see supplementary for full details on the modeling of $<r>$). Another crucial advantage of the SASS mode is that the inherent monostability of the system, guarantees that tip-trapping and or switching between states cannot occur thus providing robustness and stability. Furthermore, small drifts in the resonance curve are unlikely to affect the operation of the instrument in the SASS mode because the set-point ratio is much smaller than one $A_{sp}/A_0$<<1 (see supplementary for details). Finally, we show an experimental example of the differences between imaging in the SASS (Fig. 4(b)) and the L-state (Fig. 4(c)) modes; the only two possibilities with small values of $A_0$. These are DNA molecules on a mica surface where standard sample preparation has been followed. As predicted, there is a step change in both resolution (Fig. 4(d)) and signal-to-noise ratio when using SASS. Note that background roughness is less than half in SASS (Figs. 4(b) and 4(c)). It should be noted however, that the appearance of the N and SASS region critically depends on the use of both ultra-sharp tips and a minimum cantilever stiffness as detailed in this study. It is also worth noting that the double helix of a single dsDNA molecule has been recently[27] resolved using the SASS region in AM AFM and the same method has been employed to image DNA topoisomerases with high resolution[52]. Wastl et al. have also reported enhanced resolution in similar conditions while employing FM AFM[26, 53]. Furthermore, the authors of this work have recently shown[54] that there is a functional relationship between the sharpness of the tip, plateaus in force ranging fractions or several nm in above the region of mechanical contact, i.e. where the hydration layer is located, and the SASS region and phenomena reported here. The present work is in agreement with these recent reports and further provides evidence that stability and enhanced resolution might be related to the inhibition of instability when the tip oscillates under the hydration layer, or, in any case, in the presence of the plateaus in force reported by some[54-56].

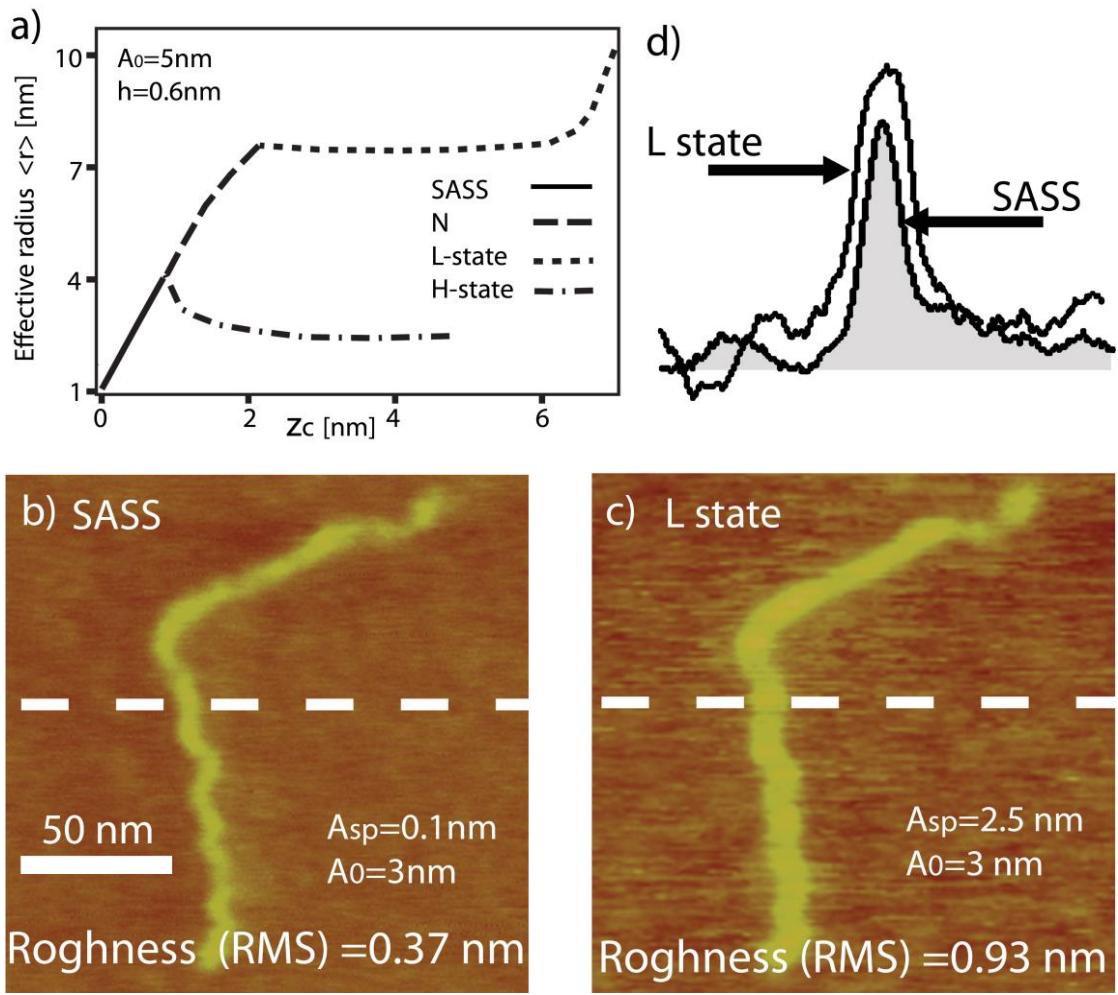

Fig 4: (a) Cantilever-sample separation ($z_c$) versus effective radius of interaction $<r>$ curve. This figure corresponds to the AD curve in Fig. 3(a). The value of $<r>$ is significantly smaller in the H state than in the L state and forms approximate plateaus; while $<r>$ is in minimized in the SASS region. Experimental examples of differences in resolution in the (b) SASS mode and the (c) L state obtained with a very sharp R<5nm tip. Background surface roughness in RMS (0.37 nm and 0.93 nm in SASS and the L state respectively) indicates that noise and is significantly decreased in the latter. (d) The width of the molecules can be compared by looking at the cross sections of these images where it can readily be observed that the width in SASS is approximately half that in the L state. Experimental parameters: $f=f_0 \sim 300$ kHz, $k \sim 40$N/m, $Q \sim 500$ and RH~40%.

In summary, the appearance of a single attractor at just several angstroms of separation is a key physical phenomenon for future developments in high resolution and low wear ambient AFM. The single attractor provides high stability to the system and the effective area of interaction is reduced there. This study should be general for dynamic interactions between hydrated surfaces with very small effective radii, i.e. surfaces interacting via hydrated nanoscale asperities or rugosities, nanoscale mechanical actuators, protein-membrane dynamic interactions, nanoparticle interactions, powders and granular materials. The results further imply that dynamic nanoscale phenomena is highly sensitive to the effective radii and the spring constant k. For example, for dynamic nanoscale oscillators the N and SASS regions are only manifest when the radii is in the order of just a few nanometers and when the spring stiffness is of at least a few units or tens of N/m. This work provides evidence of a dependence of the dynamics of nanoscale systems on the dimensions of the interacting features where only a few nm of difference might lead to the display of novel phenomena.